\pdfoutput=1
\documentclass[10pt,letterpaper,twocolumn]{article}

\usepackage[T1]{fontenc}
\usepackage[utf8]{inputenc}
\usepackage[margin=0.75in]{geometry}
\usepackage{newtxtext,newtxmath}
\usepackage{microtype}
\usepackage{graphicx}
\usepackage{amsmath}
\usepackage{booktabs,array}
\usepackage{float}
\usepackage{textcomp}
\usepackage{enumitem}
\usepackage[font=small,labelfont=bf]{caption}
\usepackage{fancyhdr}
\usepackage[hidelinks]{hyperref}
\usepackage{xurl}

\setlist[itemize]{leftmargin=*,topsep=2pt,itemsep=1pt,parsep=0pt,partopsep=0pt}

\newcommand{\Description}[1]{}
\newcommand{\authorcell}[4]{%
  \begin{minipage}[t]{0.31\textwidth}
    \centering
    {\normalsize #1\par}
    \vspace{0.12em}
    {\scriptsize #2\par}
    {\scriptsize #3\par}
    {\scriptsize \nolinkurl{#4}\par}
  \end{minipage}%
}
\newcommand{\authorcellwide}[4]{%
  \begin{minipage}[t]{0.47\textwidth}
    \centering
    {\normalsize #1\par}
    \vspace{0.12em}
    {\scriptsize #2\par}
    {\scriptsize #3\par}
    {\scriptsize \nolinkurl{#4}\par}
  \end{minipage}%
}

\newcommand{\paperpdfauthor}{JinFeng Xie, Chengfu Ou, Peipeng Yu, Xiaoyu Zhou, Dingding Huang, Jianwei Fei, Zixuan Shen, Zhihua Xia}
\newcommand{\paperkeywords}{AIGC, digital watermarking, diffusion models, tamper localization, provenance verification, latent fingerprinting, content integrity}
\newcommand{\paperabstract}{The rapid adoption of diffusion-based generative models has intensified concerns over the attribution and integrity of AI-generated content (AIGC). Existing single-domain watermarking methods either fail under regeneration, remain vulnerable to black-box reprompting that enables adversarial framing, or provide no spatial evidence for tampered regions. We propose Dual-Guard, a dual-channel latent watermarking framework for practical provenance verification, framing resistance, and region-level tamper localization. Dual-Guard combines two complementary anchors: a Gaussian Shading watermark in the initial diffusion noise as a global provenance signal, and a Latent Fingerprint Codec in the final denoised latent as a structured content anchor. Reprompting tends to preserve the former while breaking the latter, whereas localized edits disturb the content anchor only in tampered regions. In Full mode on a 2,400-sample benchmark, Dual-Guard keeps clean-image authentication false rejection and tamper false alarm below one half of one percent, while maintaining near-complete detection under reprompting, diffusion editing, and eight local tampering attacks.}

\hypersetup{
  pdftitle={Dual-Guard: Dual-Channel Latent Watermarking for Provenance and Tamper Localization in Diffusion Images},
  pdfauthor={\paperpdfauthor}
}
\fancypagestyle{firstpage}{%
  \fancyhf{}

  \fancyfoot[L]{\footnotesize Corresponding author: \nolinkurl{xia_zhihua@163.com}}
  \fancyfoot[C]{\footnotesize \thepage}
}

\begin{document}

\twocolumn[
\begin{@twocolumnfalse}
\begin{center}
{\LARGE\bfseries Dual-Guard: Dual-Channel Latent Watermarking for Provenance and Tamper Localization in Diffusion Images\par}
\vspace{0.55em}
\authorcell{JinFeng Xie}{Jinan University}{Guangzhou, China}{jinfengxie@stu2024.jnu.edu.cn}\hfill
\authorcell{Chengfu Ou}{Jinan University}{Guangzhou, China}{hahally@stu2025.jnu.edu.cn}\hfill
\authorcell{Peipeng Yu}{Jinan University}{Guangzhou, China}{ypp865@163.com}\par
\vspace{0.22em}
\authorcell{Xiaoyu Zhou}{Jinan University}{Guangzhou, China}{xiaoyuzhou68@outlook.com}\hfill
\authorcell{Dingding Huang}{Jinan University}{Guangzhou, China}{tintin@stu2024.jnu.edu.cn}\hfill
\authorcell{Jianwei Fei}{University of Florence}{Florence, Italy}{fjw826244895@163.com}\par
\vspace{0.22em}
\authorcellwide{Zixuan Shen}{Jinan University}{Guangzhou, China}{m15195901120@163.com}\hfill
\authorcellwide{Zhihua Xia}{Jinan University}{Guangzhou, China}{xia_zhihua@163.com}\par
\end{center}
\vspace{0.35em}
\end{@twocolumnfalse}
]
\thispagestyle{firstpage}

{\small
\noindent\textbf{Abstract.} \paperabstract\par
\vspace{0.35em}
\par
\vspace{0.55em}
}

\section{Introduction}
\label{sec:intro}

Recent advances in diffusion-based generative models~\cite{ho2020ddpm,rombach2022ldm,saharia2022imagen} have enabled photorealistic text-to-image synthesis at unprecedented quality. Systems such as Stable Diffusion~\cite{rombach2022ldm}, DALL-E~2~\cite{ramesh2022dalle2}, and Imagen~\cite{saharia2022imagen} now create AI-generated images at massive scale. While these capabilities unlock remarkable creative potential, they also raise two basic questions: \emph{Who generated this image?} and \emph{Has this image been tampered with?} Without reliable attribution and integrity verification, AI-generated content (AIGC) weakens trust in digital media and exposes creators to unauthorized misuse, motivating both deployment-oriented provenance standards and model-side attribution schemes~\cite{c2pa2022spec,yu2021aigc_attribution}.

\begin{figure}[t]
  \centering
  \includegraphics[width=\columnwidth]{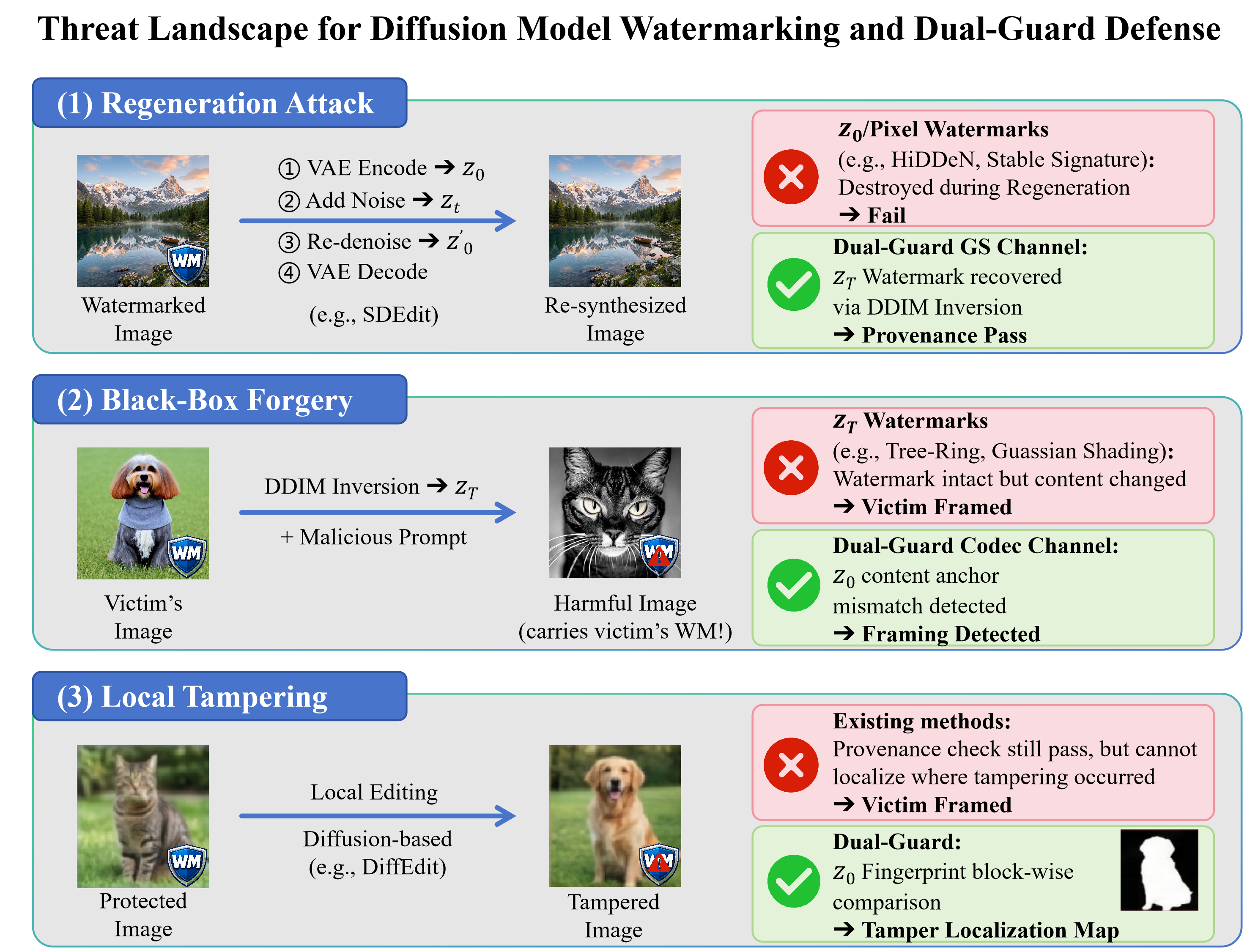}
  \caption{Threat landscape for watermarked AIGC and the Dual-Guard defense. From top to bottom, the three rows show regeneration, black-box forgery, and local tampering. In each row, the red box summarizes why a representative prior watermark family fails, while the green box shows the Dual-Guard response: Gaussian Shading (GS) preserves provenance after regeneration, the codec channel detects content-anchor mismatch under forgery, and block-wise fingerprint comparison localizes local edits.}
  \Description{Three attack scenarios against watermarked AIGC and the corresponding Dual-Guard responses.}
  \label{fig:teaser}
\end{figure}

Digital watermarking offers a principled way to embed imperceptible provenance signals into generated images. Existing methods fall into two broad paradigms~\cite{cox2024wmdiffusion}. Image-domain watermarking approaches that operate on the rendered pixels, such as HiDDeN~\cite{zhu2018hidden}, StegaStamp~\cite{tancik2020stegastamp}, and Stable Signature~\cite{fernandez2023stablesignature}, survive common distortions but remain vulnerable to \emph{regeneration attacks}, where an adversary re-encodes the image through a diffusion model and destroys the output watermark while preserving semantics~\cite{zhao2023invisible,saberi2024robustness}. Diffusion-native approaches instead watermark the initial diffusion noise, denoted $z_T$, and recover it via DDIM inversion. Yet they still provide only a \emph{global binary decision} without spatial localization, and M\"uller et al.~\cite{muller2024blackbox} show that an adversary can recover the watermark-carrying $z_T$ and reuse it with a malicious prompt, thereby \emph{framing} the legitimate owner.

Meanwhile, image tamper localization has been extensively studied in the forensics community~\cite{wu2019mantranet,guillaro2023trufor,wang2022objectformer}. These methods detect inconsistencies or learned forensic traces in manipulated photographs, but they target \emph{natural} images rather than the structured latent representations of diffusion outputs. Powerful diffusion editing tools---such as SDEdit~\cite{meng2022sdedit}, Prompt-to-Prompt~\cite{hertz2023p2p}, DiffEdit~\cite{couairon2023diffedit}, and InstructPix2Pix~\cite{brooks2023instructpix2pix}---therefore create tampering vectors that leave few classical forensic traces.

In this paper, we propose \textbf{Dual-Guard}, a unified framework that bridges the gap between robust provenance verification and region-level tamper localization for diffusion-generated images (see Figure~\ref{fig:teaser}). The key insight is to embed complementary watermark signals at \emph{two distinct positions} within the latent diffusion pipeline: the initial diffusion noise $z_T$, which carries a globally robust provenance watermark via Gaussian Shading, and the final denoised latent $z_0$ before VAE decoding, which carries a spatially structured fingerprint via a learned codec that enables block-wise tamper localization. The $z_0$ fingerprint also serves as a \emph{content anchor}: reprompting typically leads to a substantially different $z_0$, causing a marked drop in codec score even when the $z_T$ watermark is preserved. During verification, the two channels provide complementary evidence: the Gaussian Shading channel determines whether the image was generated by the protected model, while the codec channel reveals \emph{where} the image has been modified through a three-evidence fusion mechanism. Our main results focus on \emph{Full mode}, i.e., the setting where each issued image keeps an owner-side round-trip reference latent $z_0^{\text{ref}}$ for stronger block-level integrity checks.

Our main contributions are summarized as follows:
\begin{itemize}
  \item We propose \textbf{Dual-Guard}, a dual-channel latent watermarking framework that simultaneously achieves provenance verification, integrity authentication, and region-level tamper localization for diffusion-generated images. Our main experiments focus on \emph{Full mode} (with a per-image stored round-trip latent $z_0^{\text{ref}}$); a lighter reference-free path is retained only as a deployment fallback rather than the main empirical focus.
  \item We design a \textbf{Latent Fingerprint Codec} with a multi-scale gated decoder, confidence-weighted pooling, and a linear probe branch, trained with the four-stage curriculum detailed in Section~\ref{sec:training}: decoder warmup, strength scheduling, residual-budget decay, and robustness augmentation.
  \item We introduce a \textbf{three-evidence fusion} localization mechanism that combines cosine correlation, L1 deviation, and decoder BMR across latent blocks, followed by hysteresis thresholding and connected-component analysis to produce a block-level image-space heatmap while suppressing clean-image activation on our benchmark.
  \item In Full mode on a 2,400-sample benchmark, Dual-Guard achieves authentication false rejection and tamper false alarm rates of 0.3\% and 0.1\% on clean content, while reaching $\geq$99.9\% detection across reprompting, DiffEdit, and eight local tampering attacks.
\end{itemize}

\section{Related Work}
\label{sec:related}

\noindent\textbf{Watermarking for AI-Generated Images.}
Pixel-domain methods such as HiDDeN~\cite{zhu2018hidden}, StegaStamp~\cite{tancik2020stegastamp}, Distortion-Agnostic Watermarking~\cite{luo2020distortion}, and SSL-Watermark~\cite{fernandez2022sslwatermark} are robust to conventional distortions but are erased by regeneration attacks~\cite{zhao2023invisible,saberi2024robustness}. Diffusion-native approaches move the watermark into the generation process itself: Tree-Ring~\cite{wen2024treering} and RingID~\cite{ci2024ringid} encode patterns in the Fourier domain of $z_T$, Gaussian Shading~\cite{yang2024gaussianshading} uses keyed truncated-normal sampling, Stable Signature~\cite{fernandez2023stablesignature} fine-tunes the VAE decoder, and RoSteALS~\cite{bui2023rosteals}, EditGuard~\cite{zhang2024editguard}, and recent diffusion-watermark recipes~\cite{cox2024wmdiffusion} operate closer to the latent denoising path. SEAL~\cite{arabi2025seal} further conditions watermarks on semantic content via locality-sensitive hashing. Despite these advances, existing methods still return global decisions without spatial localization, and recent attack studies show that both reprompting and stronger adaptive optimization remain important threat models~\cite{muller2024blackbox,lukas2023ptw,an2024benchmarking}.

\noindent\textbf{Image Tamper Detection and Localization.}
Media forensics methods---ManTra-Net~\cite{wu2019mantranet}, CAT-Net~\cite{kwon2022catnet}, MVSS-Net~\cite{dong2022mvssnet}, ObjectFormer~\cite{wang2022objectformer}, and TruFor~\cite{guillaro2023trufor}---detect manipulations in natural photographs by exploiting compression artifacts, local anomalies, or learned noise prints. However, they are trained on traditional manipulations and do not leverage the structured latent representations of diffusion-generated images. Semantic editing via diffusion tools leaves minimal classical forensic traces, rendering these methods largely ineffective against generative tampering.

\noindent\textbf{Diffusion Editing and Security Threats.}
Recent diffusion editing tools---SDEdit~\cite{meng2022sdedit}, Prompt-to-Prompt~\cite{hertz2023p2p}, DiffEdit~\cite{couairon2023diffedit}, and InstructPix2Pix~\cite{brooks2023instructpix2pix}---enable seamless image manipulation that constitutes powerful attack vectors against watermarked AIGC. In particular, reprompting attacks~\cite{muller2024blackbox} can produce harmful content carrying the victim's watermark, while selective region editing can bypass global provenance checks. More broadly, adaptive watermark attacks highlight that robustness claims must be stress-tested beyond benign post-processing~\cite{lukas2023ptw,an2024benchmarking}. Existing diffusion watermarking methods still do not provide localized spatial evidence of where tampering occurred under a platform-side, reference-assisted verification setting---the gap that Dual-Guard targets.

\section{Methodology}
\label{sec:method}

\begin{figure*}[t]
  \centering
  \includegraphics[width=\textwidth]{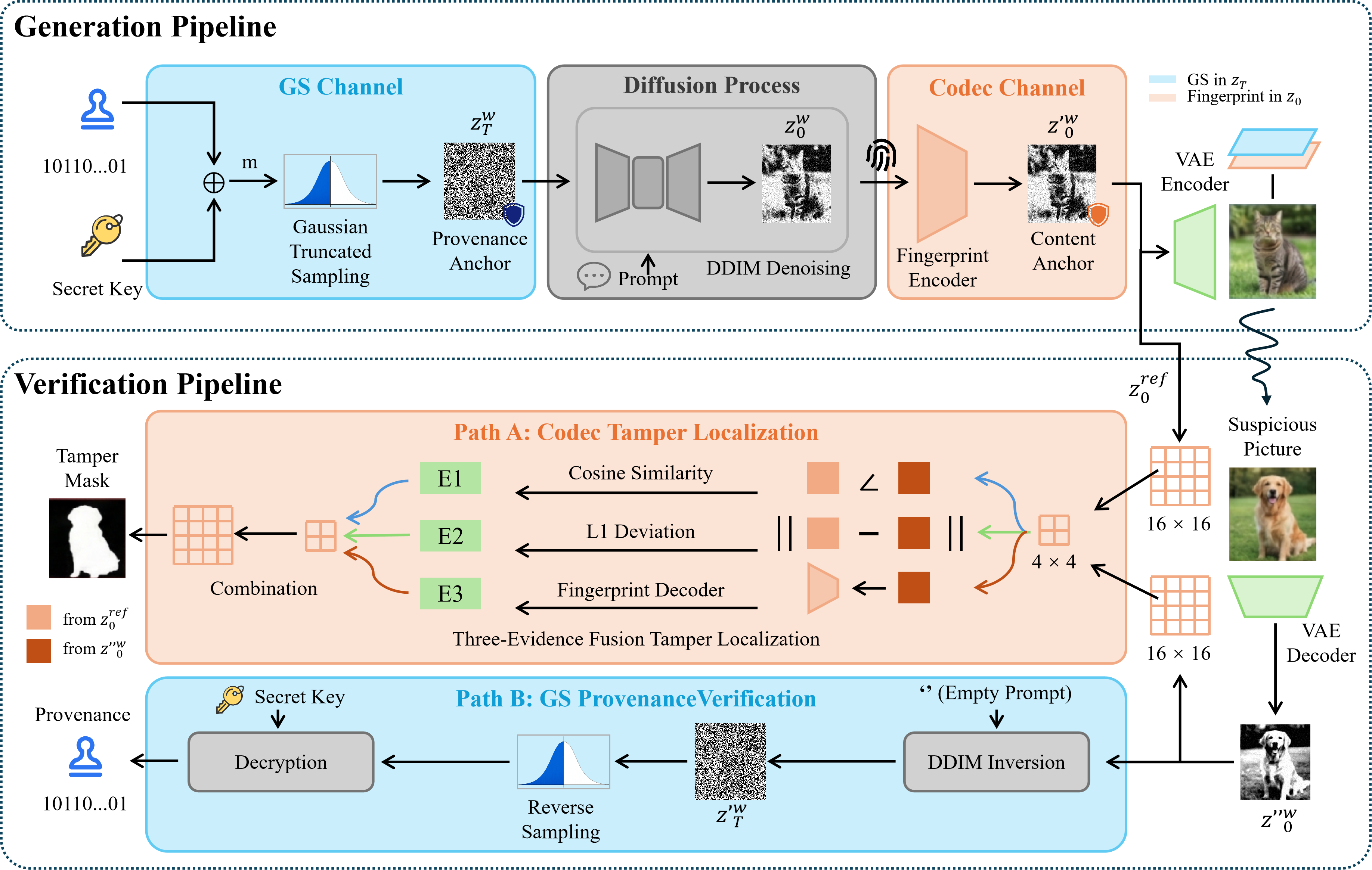}
  \caption{\textbf{Overview of the Dual-Guard framework.} The top half is the generation pipeline: the GS channel uses the registered fingerprint and secret key to sample watermarked initial noise $\tilde{z}_T$, the diffusion model denoises it to $z_0^{w}$, and the codec encoder adds a latent residual to produce the final content anchor $z_0'^{\,w}$ before VAE decoding. The bottom half is the verification pipeline. Path A re-encodes the suspicious image, compares latent blocks against the owner-side round-trip reference latent $z_0^{\text{ref}}$ stored with that image record, and converts fused cosine/L1/decoder evidence into a tamper mask. Path B DDIM-inverts the same image back to an estimate $\hat{z}_T$ of the initial noise and uses it for GS-based provenance authentication.}
  \Description{The Dual-Guard system architecture showing generation and verification pipelines with dual-channel watermarking.}
  \label{fig:framework}
\end{figure*}

\subsection{Task Definition and Threat Model}
\label{sec:threat_model}

\noindent\textbf{Threat Model.}
The \emph{content owner} controls a latent diffusion model (LDM) and embeds dual watermarks at generation time.
The \emph{attacker} is assumed to have black-box access to watermarked outputs and may re-encode images through a diffusion pipeline (\emph{regeneration}), recover $z_T$ via DDIM inversion and re-run with a malicious prompt (\emph{reprompting}~\cite{muller2024blackbox}), or apply region-level editing tools after generation.
The attacker is not assumed to know the key seed $s_k$, the derived binary mask $k$, or the registered fingerprint $f_p$. Our evaluation concentrates on black-box regeneration, reprompting, and post-generation editing; stronger white-box adaptive joint optimization with access to encoder parameters, thresholds, and reference latents is outside the scope of the present benchmark, although it remains an important direction~\cite{lukas2023ptw}.

\noindent\textbf{Owner-Side Artifacts.}
Per image, the owner registers a 128-bit key seed $s_k$ and a 64-bit fingerprint $f_p$ (24 bytes combined); at verification time, a pseudorandom generator $G$ expands $s_k$ into the per-element binary mask $k = G(s_k)$.
In \emph{Full mode}, the same registry record stores a round-trip reference latent $z_0^{\text{ref}}$ ($\sim$64~KB), obtained by VAE re-encoding the issued watermarked image once and linked to its provenance record.
A \emph{Lite mode} variant (no $z_0^{\text{ref}}$) relies on decoder BMR alone, trading localization precision for lower storage. In this submission it is treated as a deployment fallback rather than the main empirical focus.

\noindent\textbf{Registry Assumption.}
Verification assumes the platform first resolves a candidate record and retrieves $(s_k, f_p, z_0^{\text{ref}})$ from an owner-side registry or service log. In a deployment setting, this record can be tied to the provenance handle associated with the GS channel or to the corresponding issuance log; our experiments therefore study \emph{closed-set verification} of a claimed image--record pair rather than open-set fingerprint search.
In plain terms, the verifier does not search over every image ever issued. It checks whether a suspicious image matches one claimed owner-side record.

\noindent\textbf{Decision Outputs.}
The system returns a provenance verdict (authenticated or rejected), an integrity verdict (clean or tampered), and, when tampering is detected, a $16{\times}16$ block-level heatmap upsampled to image resolution. Unless noted otherwise, all main experiments report Full-mode results.

\noindent\textbf{Dual-Channel Intuition.}
The LDM pipeline exposes two critical nodes---the initial noise $z_T$ and the clean latent $z_0$---that fail under different attacks.
Dual-Guard therefore embeds fingerprints at both: the $z_T$ watermark (Gaussian Shading) acts as a \textbf{provenance anchor} that survives regeneration, while the $z_0$ fingerprint (learned Codec) acts as a \textbf{content anchor}---reprompting typically shifts $z_0$ enough to depress the codec score, whereas local edits disrupt only tampered blocks and thus enable spatial localization.

\noindent\textbf{Notation.}
Let $\mathcal{E}$ and $\mathcal{D}$ denote the VAE encoder and decoder of a latent diffusion model.
Given a text prompt $\tau$, standard generation proceeds by sampling initial Gaussian noise $z_T \sim \mathcal{N}(0, I)$, then iteratively denoising via a DDIM scheduler~\cite{song2021ddim} to obtain the clean latent $z_0$, which is decoded to the image $x = \mathcal{D}(z_0)$.
For each protected image, the owner registers a 64-bit fingerprint $f_p \in \{0,1\}^{N_{\text{fp}}}$.
The owner also samples a 128-bit key seed $s_k \in \{0,1\}^{128}$ and expands it through a pseudorandom generator into a per-element binary mask $k = G(s_k) \in \{0,1\}^{C \times H \times W}$.

\noindent\textbf{Dual-Guard Generation Pipeline.}
As illustrated in Figure~\ref{fig:framework}, Dual-Guard embeds $f_p$ at two positions. In the GS channel, the initial noise $z_T$ is replaced by a watermarked substitute $\tilde{z}_T$ generated via keyed truncated-normal sampling (Section~\ref{sec:gs_channel}). In the codec channel, after denoising, the clean latent $z_0$ is modified by a learned encoder to produce $\tilde{z}_0 = z_0 + \alpha \cdot E(z_0, f_p)$, which is then decoded to the watermarked image (Section~\ref{sec:codec}).

\subsection{Gaussian Shading Channel}
\label{sec:gs_channel}

The GS channel provides \emph{global} provenance: it answers \emph{``Was this image generated by the protected model?''} robustly against regeneration attacks.

\noindent\textbf{Watermarked Noise Generation.}
Given a binary payload $f_p \in \{0,1\}^{N_{\text{fp}}}$ and a binary mask $k = G(s_k) \in \{0,1\}^{C \times H \times W}$ expanded from the seed $s_k$, we first tile $f_p$ to match the latent spatial dimension and compute a watermark mask $m = (f_p \oplus k) \in \{0,1\}^{C \times H \times W}$ via bitwise XOR.
Each element of $\tilde{z}_T$ is then drawn from a truncated normal distribution conditioned on the corresponding bit of $m$:
\begin{equation}
  [\tilde{z}_T]_i \sim
  \begin{cases}
    \mathcal{TN}(-\infty, 0) & \text{if } m_i = 0 \\
    \mathcal{TN}(0, +\infty)  & \text{if } m_i = 1
  \end{cases}
\end{equation}
where $\mathcal{TN}(a,b)$ denotes a standard normal distribution truncated to $(a, b)$.
Under the GS sampling construction, each element preserves the same standard-Gaussian marginal as ordinary diffusion noise. Consistent with the original GS analysis~\cite{yang2024gaussianshading}, this channel is therefore expected to add little extra perceptual distortion.

\noindent\textbf{Extraction via DDIM Inversion.}
Given a suspicious image $\hat{x}$, we first obtain $\hat{z}_0 = \mathcal{E}(\hat{x})$ and apply deterministic DDIM inversion~\cite{song2021ddim} to recover an estimate $\hat{z}_T$.
We then binarize: $\hat{m}_i = \mathbf{1}[\hat{z}_{T,i} > 0]$, and recover the tiled payload map $\hat{F} = \hat{m} \oplus k$.
Let $\pi(j)$ denote the index set mapped to the $j$-th payload bit during tiling. We collapse the tiled map back to the 64-bit payload by majority voting over each set $\pi(j)$, and define $\text{BMR}_{\text{GS}}$ as the fraction of recovered bits matching the registered fingerprint $f_p$.
The released threshold profile uses $\tau_{\text{GS}} = 0.995$ for provenance authentication.
Under standard Gaussian noise, the expected BMR is 0.5; the threshold is selected once on the held-out calibration split described in Section~\ref{sec:setup}, leaving a wide margin to chance-level negatives.

\subsection{Latent Fingerprint Codec}
\label{sec:codec}

The Codec channel provides \emph{spatial} evidence: it identifies \emph{which regions} of the image have been modified, enabling region-level tamper localization. Each protected image is associated with a registered 64-bit fingerprint $f_p$; this paper evaluates the stage where that record has already been retrieved, i.e., the verifier checks one claimed record rather than searching the whole registry.

\noindent\textbf{Encoder: Residual Embedding.}
The encoder $E$ maps the clean latent $z_0 \in \mathbb{R}^{4 \times 64 \times 64}$ and fingerprint $f_p$ to a learned residual $r \in \mathbb{R}^{4 \times 64 \times 64}$. Concretely, a \texttt{FingerprintExpander} projects $f_p$ through a linear layer, GroupNorm, and LeakyReLU into a spatial feature map of shape $(4,64,64)$; this map is concatenated with $z_0$ and processed by three residual blocks to produce a raw residual $\hat{r}$. We clamp $\hat{r}$ to $[-2,2]$ and form the watermarked latent as $\tilde{z}_0 = z_0 + \alpha r$ with fixed deployment strength $\alpha = 0.3$.
Zero initialization of the output convolution ensures the embedding starts from a near-identity mapping, avoiding catastrophic disruption of the host latent.

\begin{figure}[t]
  \centering
  \includegraphics[width=\columnwidth]{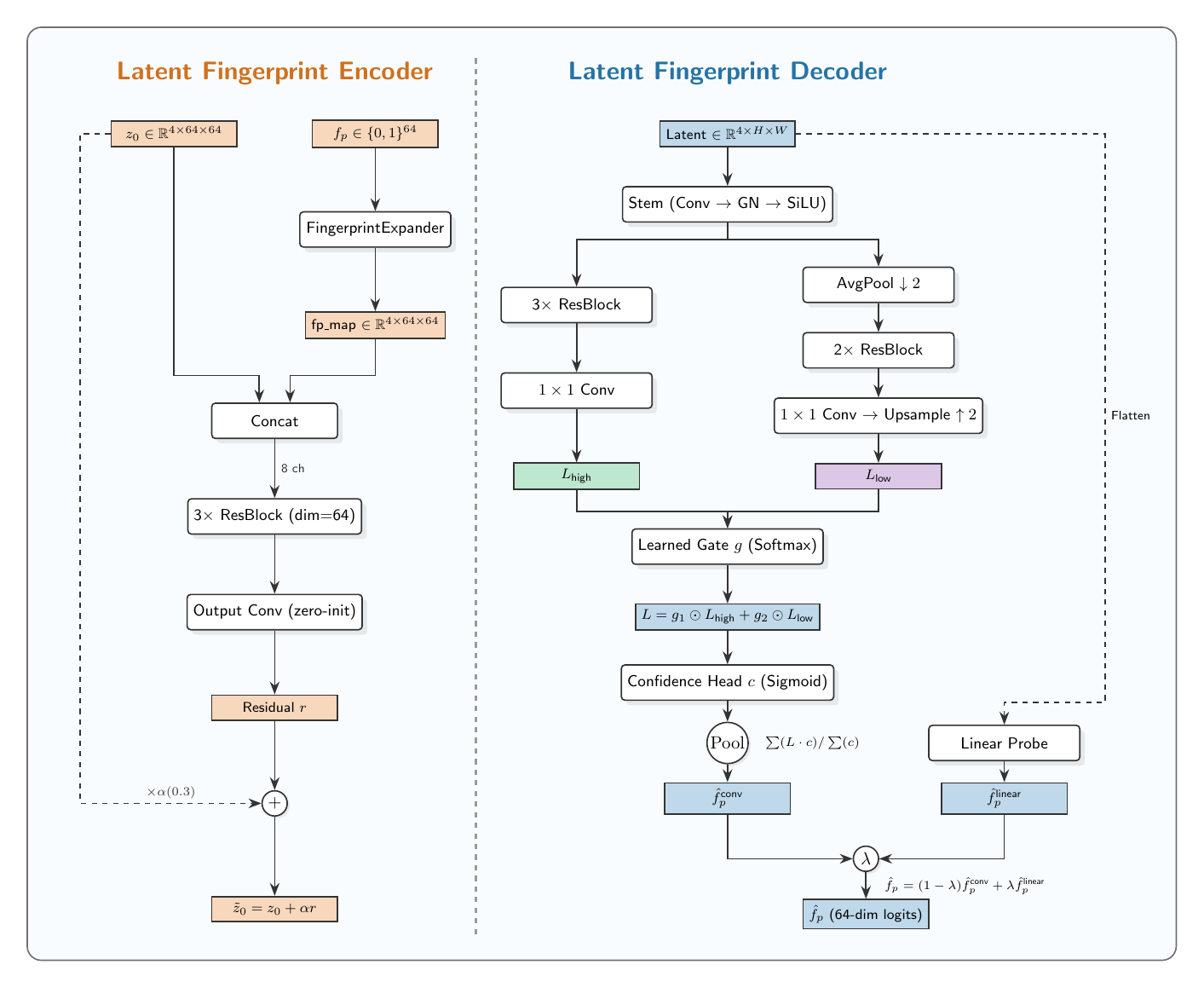}
  \caption{\textbf{Architecture of the Latent Fingerprint Codec.} Left: the encoder expands the 64-bit fingerprint to a spatial map, concatenates it with $z_0$, predicts a residual through three ResBlocks, and adds the scaled residual back to form the watermarked latent $\tilde{z}_0$. Right: the decoder first applies a shared stem, then splits into a high-resolution branch for local block cues and a low-resolution branch for context robustness. Their logits are fused by a learned gate, pooled with confidence weights, and finally combined with a linear probe so that the same decoder can operate on both full images and small latent blocks.}
  \Description{Encoder and Decoder architecture of the Latent Fingerprint Codec.}
  \label{fig:codec}
\end{figure}

\noindent\textbf{Decoder: Multi-Scale Gated Fusion.}
A key challenge for tamper localization is that block-level decoding operates on $4{\times}4$ patches---far smaller than the full $64{\times}64$ training inputs.
To handle this, we design a \textbf{Decoder} (Figure~\ref{fig:codec}), a resolution-agnostic architecture.
The input latent first passes through a shared \emph{stem} layer (3$\times$3 Conv $\to$ GroupNorm $\to$ SiLU) to project from 4 channels to hidden dimension $d=64$, and then branches into a high-resolution path that preserves local detail and a low-resolution path that captures broader context for robustness. Each branch predicts per-pixel fingerprint logits. Their features are fused by a learned gate, pooled by a confidence head, and combined with a lightweight linear probe operating on the flattened latent. This design lets the decoder mix local spatial evidence with global spread-spectrum evidence instead of relying on a single scale.

\noindent\textbf{Error Correction Coding (ECC).}
A $3\times$ repetition code expands the 64-bit payload into a 192-bit codeword~\cite{lin2004ecc}. During extraction, the decoder first predicts the 192 coded bits, then applies majority voting within each triple to recover the 64-bit payload; all reported codec BMR values are computed on the decoded 64-bit payload.

\subsection{Curriculum Learning Training Strategy}
\label{sec:training}

Training a codec that must be simultaneously imperceptible (low distortion to $z_0$) and robust (extractable even from $4{\times}4$ blocks after VAE \mbox{encode-decode} roundtrip) poses contradictory objectives.
Our default run uses AdamW with OneCycleLR (max learning rate $3{\times}10^{-4}$, weight decay $10^{-4}$), batch size 16, and 4,000 optimization steps. Training samples come from a 512-sample latent bank built with the same SD~2.1-base / 50-step DDIM / guidance-7.5 pipeline as the main experiments; for each prompt we store both the clean $z_0$ latent and its VAE round-trip latent so that the decoder already sees verification-time distribution shift during training. Following the spirit of curriculum learning~\cite{bengio2009curriculum}, we resolve the fidelity--robustness tension with four coupled schedules:

\noindent\textbf{Phase 1 --- Decoder Warmup (220 steps).}
The encoder is frozen and replaced by a fixed spread-spectrum embedder: the coded fingerprint is injected into $z_0$ with normalized random carrier vectors that are sampled once and held fixed during warmup, with embedding strength sampled in $[0.22, 0.65]$.
This stage lets the decoder learn to read a dense analytic watermark \emph{before} joint encoder--decoder optimization begins.

\noindent\textbf{Phase 2 --- Strength Curriculum (1,000-step ramp).}
Once joint training starts, the neural encoder is activated and the embedding strength is annealed from $\alpha=1.2$ down to $\alpha=0.35$.
The decoder is trained with a bitwise BCE loss, while the encoder is regularized by a latent-domain structural penalty based on SSIM~\cite{wang2004ssim} to preserve fidelity.
This schedule is used only during training; the released inference path uses the fixed encoder strength $\alpha = 0.3$ described above.

\noindent\textbf{Phase 3 --- Residual Budget Decay (1,600-step ramp).}
A hinge-style penalty suppresses per-element residual magnitudes that exceed a shrinking budget $\beta_t$, decaying from 0.20 to 0.08.
This encourages the encoder to concentrate embedding energy on the latent locations that are most informative for the decoder.

\noindent\textbf{Phase 4 --- Robustness Augmentation.}\par
\noindent Latent-space augmentations (Gaussian noise, blur, dropout/cutout, and quantization) ramp in over 600 steps.
During the final third of training, we additionally enable a periodic VAE encode--decode cycle (every 3 steps) plus image-space resize/noise/blur perturbations ramped over 900 steps, so the decoder learns the inference-time distribution shift.

\subsection{Three-Evidence Fusion Tamper Localization}
\label{sec:localization}

Given a suspicious image $\hat{x}$, we obtain its latent $\hat{z}_0 = \mathcal{E}(\hat{x})$ and compare it against a \emph{reference latent} $z_0^{\text{ref}}$ retrieved from the same owner-side record as the provenance handle for the protected image.
We partition $\hat{z}_0$ and $z_0^{\text{ref}}$ into a $16{\times}16$ grid of non-overlapping $4{\times}4$ blocks (matching the $64{\times}64$ latent resolution of SD 2.1). For each block $(i,j)$, we compute three evidence scores. The first is cosine correlation, re-mapped to $[0,1]$ via $(\cos + 1)/2$; tampered blocks typically depress this score. The second is normalized L1 deviation over the $4\times4\times4$ block, which captures absolute latent displacement relative to the reference.

\noindent\textbf{Evidence 3 --- Decoder BMR.}
Each $4{\times}4$ block is fed to the Codec decoder to predict the 192 coded bits, which are then repetition-decoded back to 64 payload bits before computing the block-wise Bit Match Rate $\text{BMR}_{ij}$.
A block with intact fingerprint content yields $\text{BMR}_{ij} \approx 1.0$; a tampered block yields $\text{BMR}_{ij} \approx 0.5$ (chance level).
In practice, this quantity is best interpreted as a \emph{local fingerprint-consistency score} on the repeated coded payload rather than as recovery of a fully self-contained 64-bit message from each tiny latent block. Note that this evidence is \emph{reference-free}: it requires only the known fingerprint $f_p$, not the stored $z_0^{\text{ref}}$.

\begin{sloppypar}
\noindent\textbf{Fusion and Scoring.}
The three evidences are fused into a per-block consistency score:
\begin{equation}
  Q_{ij} = w_1\,\text{Corr}_{ij} + w_2\,\tilde{\Delta}_{ij} + w_3\,\text{BMR}_{ij},
\end{equation}
where $\text{Corr}_{ij} \in [0,1]$ is the re-mapped cosine similarity, $\tilde{\Delta}_{ij}=1-\mathrm{clip}(\lVert \Delta_{ij}\rVert_1/s,0,1)$ is the normalized L1-based similarity with the adaptive scalar $s = \mu_{L1} + 2\sigma_{L1}$ computed from the per-block L1 statistics of the current image, and higher $Q_{ij}$ means that the block looks more authentic.
Weights $(w_1, w_2, w_3){=}(0.50, 0.30, 0.20)$ are calibrated on 220 authentic--tampered pairs.
When $z_0^{\text{ref}}$ is unavailable, the system falls back to Evidence~3 alone ($Q_{ij} {=} \text{BMR}_{ij}$), requiring only $f_p$.

\noindent\textbf{Post-Processing.}
Blocks with $Q_{ij} < \tau_{\text{cand}}$ are treated as tamper candidates, where $\tau_{\text{cand}}=\tau_{\text{codec}}=0.8919$ in the released profile.
A stricter seed threshold $\tau_{\text{seed}}=\tau_{\text{cand}}-0.015$ retains only components touching a strong-mismatch core.
We require at least two 8-neighbors, suppress components smaller than 6 blocks, and prune the highest-score 10\% fringe blocks inside each retained component (prune quantile 0.90).
The resulting mask is summarized by a tamper-area ratio $\rho$; the default profile raises an integrity alarm when $\rho > 0.0117$ and the global codec consistency also drops below $\tau_{\text{codec}} + 0.01$, with a conservative large-component override for strongly mismatched regions.
\end{sloppypar}

\noindent\textbf{End-to-End Workflow (Full Mode).}
The algorithm flow below summarizes the same story in a compact review-friendly form: issuance first creates the owner-side record, and verification then consumes that record before running provenance, content, and localization checks.
\begin{center}
\footnotesize
\setlength{\tabcolsep}{3pt}
\renewcommand{\arraystretch}{1.08}
\begin{tabular}{@{}p{0.08\columnwidth}p{0.22\columnwidth}p{0.60\columnwidth}@{}}
\toprule
\textbf{Step} & \textbf{Stage} & \textbf{Operation and Outcome} \\
\midrule
1 & Record setup & Register key seed $s_k$ and fingerprint $f_p$ for the issued image. \\
2 & GS issuance & Sample watermarked initial noise $\tilde{z}_T$ and denoise it to the clean latent $z_0$. \\
3 & Codec issuance & Write the latent fingerprint, decode the protected image $x^{w}$, and in Full mode re-encode $x^{w}$ once to store the round-trip latent $z_0^{\text{ref}}$ in the owner-side record. \\
4 & Record lookup & Use the provenance handle or issuance log to retrieve the claimed record $(s_k,f_p,z_0^{\text{ref}})$. \\
5 & Provenance gate & DDIM-invert $\hat{x}$ to $\hat{z}_T$ and test the GS watermark. Failure rejects the claim immediately. \\
6 & Content gate & Re-encode $\hat{x}$ to $\hat{z}_0$ and test global codec consistency. Failure indicates a content mismatch, as in reprompting. \\
7 & Localization & Only if both gates pass, compare blocks in $\hat{z}_0$ against $z_0^{\text{ref}}$, fuse Corr/L1/BMR evidence, and output the tamper verdict plus heatmap. \\
\bottomrule
\end{tabular}
\end{center}
This matches Figure~\ref{fig:framework}: Steps 1--3 correspond to issuance, Step 4 resolves the owner-side record, and Steps 5--7 make the GS gate, codec gate, and final localization order explicit.

\section{Experiments}
\label{sec:experiments}

\subsection{Experimental Setup}
\label{sec:setup}

Unless otherwise stated, all experiments use Stable Diffusion 2.1-base at $512 \times 512$ resolution with 50-step DDIM sampling and classifier-free guidance scale 7.5. The GS branch carries a 64-bit provenance payload via keyed truncated-normal sampling, while the codec branch embeds a 192-bit latent fingerprint (64 payload bits $\times$ 3 repetition code) and supplies block-wise tamper evidence. In Full mode, the owner-side record linked to the provenance handle also stores a round-trip reference latent: the watermarked image is re-encoded via the VAE encoder to produce $z_0^{\text{ref}}$, matching the distribution seen during verification. The 512-sample latent bank is used only for codec training, whereas threshold calibration and all reported tables use separate held-out experiment splits. All thresholds and fusion hyperparameters are calibrated once on a held-out split of 220 authentic and 220 tampered image pairs and then frozen. The released profile uses $\tau_{\text{GS}}{=}0.995$, $\tau_{\text{codec}}{=}0.8919$, tamper-area threshold $\rho_{\text{tamper}}{=}0.0117$, global-guard margin 0.01, 0.015 hysteresis margin, 2-neighbor retention, 6-block minimum components, prune quantile 0.90, and 0.50/0.30/0.20 fusion weights. Unless explicitly noted, all tables correspond to Full mode.

To reduce prompt-selection bias and stay close to recent shared-prompt watermark robustness evaluations~\cite{an2024benchmarking}, we freeze a 1,000-prompt snapshot from Gustavosta\footnote{\url{https://huggingface.co/datasets/Gustavosta/Stable-Diffusion-Prompts}} for all experiments. Our protocol covers four settings: \textit{original} (clean), \textit{reprompt} (black-box regeneration using reused or recovered initial noise under a different prompt), \textit{local} (eight spatial tampering operators applied only inside sampled edit regions: black masking, noise, blur, JPEG, pixelation, copy-move, color shift, and text overlay; on the final block-level GT masks, the average tampered area is 13.3\%, ranging from 12.4\% to 14.0\%), and \textit{DiffEdit} (DDIM inversion plus prompt editing with guidance 5.0 and switch step 24). Wilson 95\% CIs are reported for all binomial rates. Unless otherwise specified, all results correspond to Full mode. Table~\ref{tab:exp_overview} summarizes the evaluation modules.

\begin{table}[t]
  \caption{Overview of experimental modules. All main experiments use the same frozen 1,000-prompt snapshot (SD 2.1-base, $512{\times}512$, 50-step DDIM, guidance 7.5).}
  \label{tab:exp_overview}
  \centering
  \resizebox{\columnwidth}{!}{
  \begin{tabular}{@{}llll@{}}
    \toprule
    Module & Samples & Primary Metrics & Purpose \\
    \midrule
    Provenance      & 4$\times$1\,k groups      & Auth, Prec, Rec, AUC       & Threshold stability \\
    Attack          & 4$\times$2.4\,k splits    & Reject, Detect, Loc.\ gain & Branch complementarity \\
    Localization    & 1\,k tamp.+1\,k clean     & IoU, F1, Prec, Rec         & Spatial precision \\
    Image quality   & 1\,k paired              & PSNR, SSIM, LPIPS, CLIP    & Fidelity impact \\
    Ext.\ baselines & Same prompt snapshot      & Native metrics             & Cross-method comparison \\
    \bottomrule
  \end{tabular}}
\end{table}

\subsection{Provenance Verification}
\label{sec:provenance}


Table~\ref{tab:provenance_main} reports provenance results on a 4,000-image closed-set suite (1,000 authentic vs.\ 1,000 each for plain SD, GS-only, and codec-only). The latter three groups are controlled unauthorized-content negatives that isolate each gate rather than an exhaustive open-world benchmark, so this table should be read as claimed image-record verification rather than open-world attribution. Dual-Guard achieves 0.999 authentication pass on authentic images (Wilson 95\% CI [0.994, 1.000]) with precision 1.000 and AUC 1.000. Neither single-channel group passes, confirming the need for the dual pipeline.

\begin{table}[t]
  \caption{Provenance verification on the 4,000-image suite (1,000 per group). Dual-Guard achieves near-perfect discrimination (AUC 1.000) between authentic and unauthorized content.}
  \label{tab:provenance_main}
  \centering
  \resizebox{\columnwidth}{!}{
  \begin{tabular}{lccccccc}
    \toprule
    Group & Count & \shortstack{Auth\\Pass} & \shortstack{GS\\Pass} & \shortstack{GS Extr.\\Rate} & \shortstack{Codec Extr.\\Rate} & \shortstack{Combined\\Extr.\ Rate} & \shortstack{Auth\\CI$_{95}$} \\
    \midrule
    Authentic  & 1000 & 0.999 & 1.000 & $1.000{\pm}0.001$ & $0.939{\pm}0.028$ & $0.982{\pm}0.009$ & [0.994, 1.000] \\
    Plain SD   & 1000 & 0.000 & 0.000 & $0.501{\pm}0.030$ & $0.500{\pm}0.035$ & $0.501{\pm}0.023$ & [0.000, 0.004] \\
    GS-only    & 1000 & 0.000 & 1.000 & $1.000{\pm}0.000$ & $0.500{\pm}0.036$ & $0.850{\pm}0.011$ & [0.000, 0.004] \\
    Codec-only & 1000 & 0.000 & 0.000 & $0.500{\pm}0.031$ & $0.940{\pm}0.028$ & $0.632{\pm}0.023$ & [0.000, 0.004] \\
    \midrule
    \multicolumn{8}{l}{\emph{Overall}: TP=999, TN=3000, FP=0, FN=1 $\to$ Precision=1.000, Recall=0.999, F1=0.999, AUC=1.000, TPR@1\%FPR=1.000} \\
    \bottomrule
  \end{tabular}}
\end{table}

\subsection{Attack Resilience}
\label{sec:attacks}


Table~\ref{tab:attack_main} presents the main robustness benchmark (2,400 samples per split). Dual-Guard achieves 1.000 rejection for reprompt and DiffEdit (lower bound 0.998) and 0.999 for local tampering (lower bound 0.997); on clean images, authentication false rejection and tamper false alarm are only 0.003 and 0.001. Localization gain reaches 97.5\% for local attacks and 60.1\% for DiffEdit, confirming that neither channel alone is sufficient.

\begin{table}[t]
  \caption{Attack resilience on the 2,400-sample benchmark. \emph{Auth reject} is the final non-authenticated rate; \emph{tamper detect} is the integrity-alarm rate. Under attack, lower \emph{global dual pass} is better, and \emph{localization gain} measures when the localization branch provides the decisive rejection.}
  \label{tab:attack_main}
  \centering
  \resizebox{\columnwidth}{!}{
  \begin{tabular}{lcccccccc}
    \toprule
    Scenario & $n$ & \shortstack{Auth\\Reject $\uparrow$} & \shortstack{Reject\\CI$_{95}$} & \shortstack{Tamper\\Detect $\uparrow$} & \shortstack{GS\\Pass} & \shortstack{Dual\\Pass $\downarrow$} & \shortstack{Loc.\\Gain $\uparrow$} & \shortstack{Codec\\Extr.} \\
    \midrule
    Original (clean) & 2400 & 0.003 & [0.002, 0.007] & 0.001 & 0.998 & 0.998 & -- & 0.939 \\
    Reprompt     & 2400 & 1.000 & [0.998, 1.000] & 1.000 & 1.000 & 0.000 & 0.000 & 0.500 \\
    Local tamper & 2400 & 0.999 & [0.997, 1.000] & 0.999 & 0.995 & 0.975 & 0.975 & 0.923 \\
    DiffEdit     & 2400 & 1.000 & [0.998, 1.000] & 1.000 & 0.633 & 0.601 & 0.601 & 0.929 \\
    \bottomrule
  \end{tabular}}
\end{table}

Table~\ref{tab:attack_local_breakdown} breaks down the eight local operators (300 samples each). All achieve 1.000 detection except text overlay (0.993). Localization gain ranges from 0.943 (noise) to 0.990 (color shift). The lower recall for text overlay (0.471 in Table~\ref{tab:localization_by_method}) reflects a structural limitation: text characters typically alter only a few thin strokes within each $32{\times}32$ pixel block, and the VAE encoder smooths these high-frequency, small-area perturbations, producing latent-space deviations that stay below the candidate threshold in many affected blocks.

\begin{table}[t]
  \caption{Per-method local tampering results (2,400 total; 300 per method). Auth rejection is 1.000 for all methods except text overlay (0.993).}
  \label{tab:attack_local_breakdown}
  \centering
  \resizebox{\columnwidth}{!}{
  \begin{tabular}{lcccccc}
    \toprule
    Method & $n$ & \shortstack{Tamper\\Detect $\uparrow$} & \shortstack{Auth\\Reject $\uparrow$} & \shortstack{GS\\Pass} & \shortstack{Dual\\Pass $\downarrow$} & \shortstack{Loc.\\Gain $\uparrow$} \\
    \midrule
    Black patch   & 300 & 1.000 & 1.000 & 1.000 & 0.960 & 0.960 \\
    Noise         & 300 & 1.000 & 1.000 & 0.987 & 0.943 & 0.943 \\
    Blur          & 300 & 1.000 & 1.000 & 0.993 & 0.967 & 0.967 \\
    JPEG          & 300 & 1.000 & 1.000 & 0.997 & 0.987 & 0.987 \\
    Pixelate      & 300 & 1.000 & 1.000 & 1.000 & 0.980 & 0.980 \\
    Copy-move     & 300 & 1.000 & 1.000 & 0.993 & 0.983 & 0.983 \\
    Color shift   & 300 & 1.000 & 1.000 & 0.997 & 0.990 & 0.990 \\
    Text overlay  & 300 & 0.993 & 0.993 & 0.993 & 0.987 & 0.987 \\
    \bottomrule
  \end{tabular}}
\end{table}

\begin{figure*}[!t]
  \centering
  \includegraphics[width=0.305\textwidth]{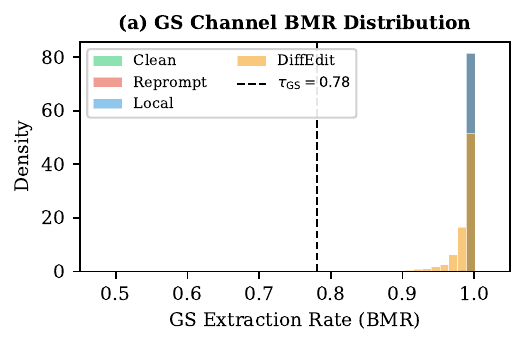}%
  \hfill
  \includegraphics[width=0.305\textwidth]{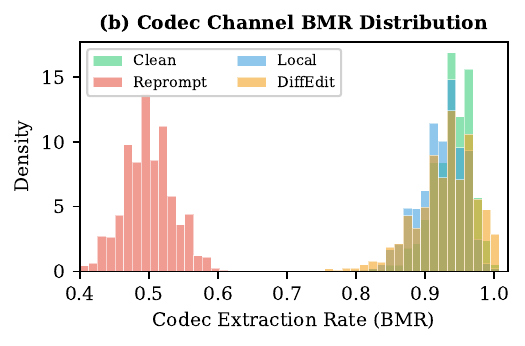}%
  \hfill
  \includegraphics[width=0.305\textwidth]{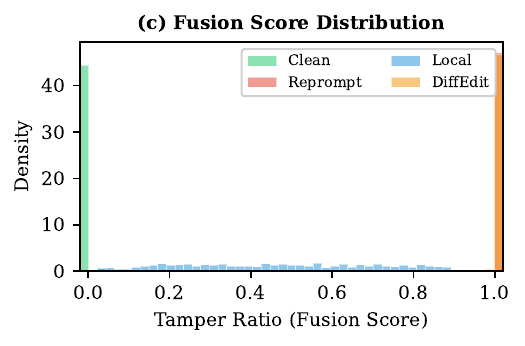}
  \caption{\textbf{Per-channel score distributions across all 2,400$\times$4 samples.} (a) GS extraction remains concentrated near 1.0 for clean images and all three attack groups, showing that the GS channel mainly answers provenance rather than integrity. (b) Codec extraction cleanly separates reprompting, which collapses to chance-level decoding around 0.5, from clean, local-tamper, and most DiffEdit samples, which stay much higher. (c) The fused tamper score turns these complementary behaviors into the final decision space: clean images cluster near 0, reprompt and DiffEdit cluster near 1, and local tampering occupies the intermediate region because only part of the image is altered.}
  \Description{Histograms of GS BMR, Codec BMR, and Fusion Score across clean, reprompt, local, and DiffEdit scenarios.}
  \label{fig:distributions}
\end{figure*}

\subsection{Tamper Localization Performance}
\label{sec:localization_results}


Table~\ref{tab:localization_main} evaluates spatial precision on 1,000 tampered and 1,000 clean images. Dual-Guard achieves 100\% image-level detection with mean IoU $0.255{\pm}0.124$ and F1 $0.392{\pm}0.145$ (recall $0.934{\pm}0.185$). The lower precision (0.273) reflects a deliberate high-recall design at $16{\times}16$ block resolution, since missing a tampered region is costlier than over-flagging in forensic use. Table~\ref{tab:localization_by_method} shows per-attack breakdowns, with copy-move (IoU 0.379) and black patch (0.335) performing best.

\begin{table}[t]
  \caption{Block-level tamper localization on 1,000 tampered and 1,000 clean images. The system operates at $16{\times}16$ block resolution ($32{\times}32$ pixels each), favoring high recall over precision.}
  \label{tab:localization_main}
  \centering
  \resizebox{\columnwidth}{!}{
  \begin{tabular}{lcccccccc}
    \toprule
    Split & $n$ & \shortstack{Img-lvl\\Detect} & IoU & F1 & Prec. & Recall & \shortstack{GT\\Area} & \shortstack{Pred.\\Area} \\
    \midrule
    Tampered & 1000 & 1.000 & $0.255{\pm}0.124$ & $0.392{\pm}0.145$ & $0.273{\pm}0.140$ & $0.934{\pm}0.185$ & 0.133 & 0.535 \\
    Clean    & 1000 & 0.000 & --    & --    & --    & --    & -- & 0.011 \\
    \bottomrule
  \end{tabular}}
\end{table}

\begin{table}[t]
  \caption{Localization metrics per attack method (125 samples each). All methods achieve 100\% image-level detection.}
  \label{tab:localization_by_method}
  \centering
  \resizebox{\columnwidth}{!}{
  \begin{tabular}{lccccccc}
    \toprule
    Method & $n$ & IoU & F1 & Prec. & Recall & \shortstack{GT\\Area} & \shortstack{Pred.\\Area} \\
    \midrule
    Black patch   & 125 & 0.335 & 0.486 & 0.335 & 1.000 & 0.124 & 0.409 \\
    Noise         & 125 & 0.177 & 0.296 & 0.177 & 1.000 & 0.140 & 0.791 \\
    Blur          & 125 & 0.212 & 0.342 & 0.212 & 1.000 & 0.131 & 0.644 \\
    JPEG          & 125 & 0.259 & 0.399 & 0.259 & 1.000 & 0.128 & 0.546 \\
    Pixelate      & 125 & 0.228 & 0.362 & 0.228 & 1.000 & 0.134 & 0.636 \\
    Copy-move     & 125 & 0.379 & 0.533 & 0.379 & 1.000 & 0.136 & 0.396 \\
    Color shift   & 125 & 0.232 & 0.364 & 0.232 & 1.000 & 0.133 & 0.619 \\
    Text overlay  & 125 & 0.216 & 0.351 & 0.361 & 0.471 & 0.140 & 0.237 \\
    \midrule
    \textbf{Average} & 1000 & \textbf{0.255} & \textbf{0.392} & \textbf{0.273} & \textbf{0.934} & 0.133 & 0.535 \\
    \bottomrule
  \end{tabular}}
\end{table}

Table~\ref{tab:localization_comparison} compares with SEAL~\cite{arabi2025seal} under the same prompt snapshot. Dual-Guard improves over SEAL by $7.1\times$ in IoU (0.255 vs.\ 0.036) and $5.8\times$ in F1 (0.392 vs.\ 0.068), while reducing clean false-alarm area from 0.050 to 0.011.

\begin{figure}[t]
  \centering
  \includegraphics[width=\columnwidth]{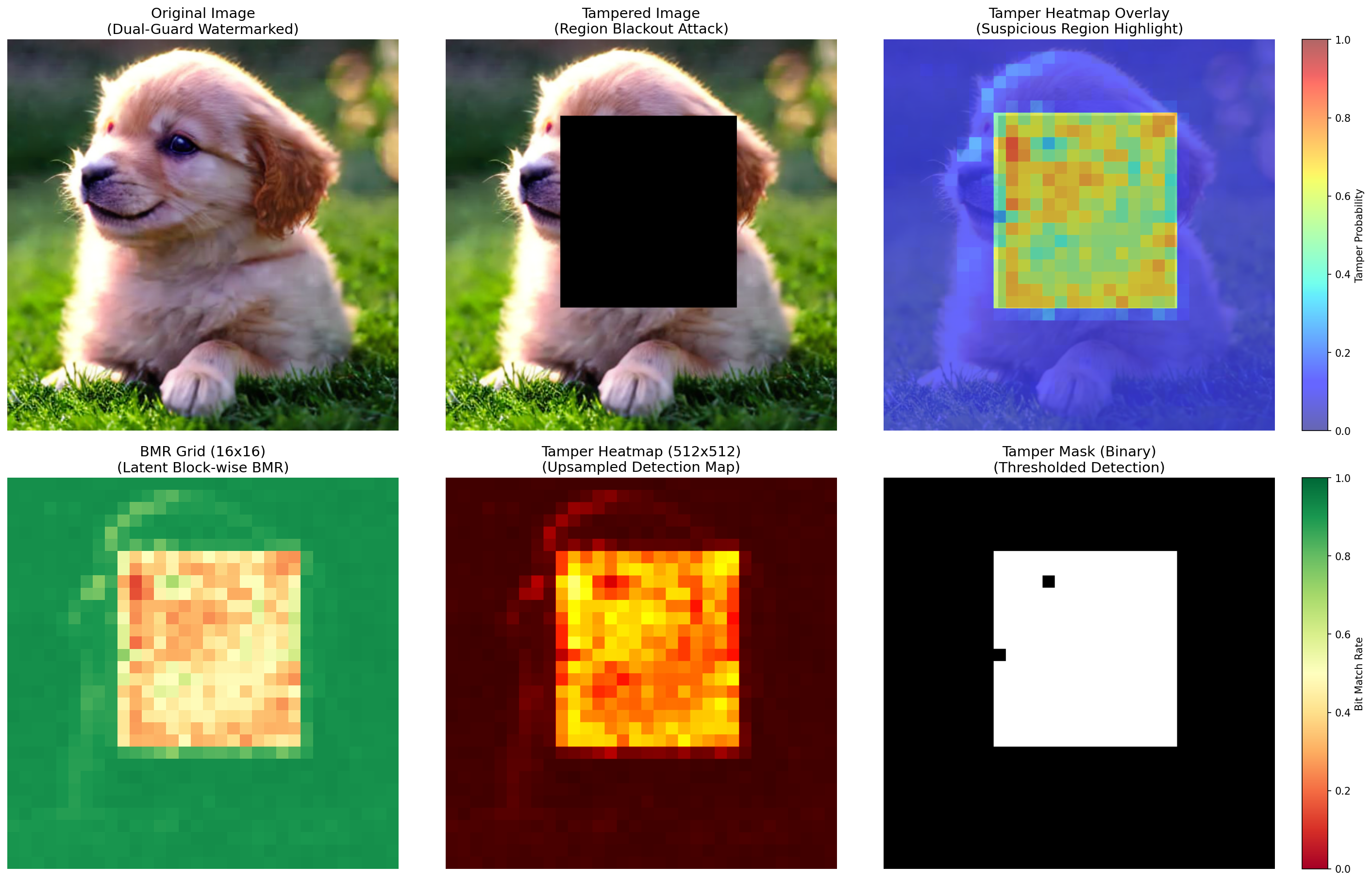}
  \caption{\textbf{One complete tamper-localization example.} Top row, from left to right: the original watermarked image, the attacked image after region blackout, and the heatmap overlaid on the suspicious image, where warmer colors mark more suspicious blocks. Bottom row: the underlying $16{\times}16$ latent-block BMR grid, the same scores upsampled to a $512{\times}512$ heatmap, and the final binary tamper mask after post-processing. The figure illustrates how Dual-Guard turns block-level evidence into an image-space localization result that roughly covers the edited square while allowing slight expansion around boundaries for higher recall.}
  \Description{Tamper detection visualization examples.}
  \label{fig:tamper_results}
\end{figure}

\begin{table}[t]
  \caption{Localization comparison with SEAL under the same 1,000-prompt snapshot.}
  \label{tab:localization_comparison}
  \centering
  \resizebox{\columnwidth}{!}{
  \begin{tabular}{lccccccc}
    \toprule
    Method & \shortstack{Img-lvl\\Detect} & IoU & F1 & Prec. & Recall & \shortstack{Clean\\Area} & \shortstack{Ratio\\vs.\ SEAL} \\
    \midrule
    Dual-Guard & 1.000 & \textbf{0.255} & \textbf{0.392} & \textbf{0.273} & \textbf{0.934} & \textbf{0.011} & -- \\
    SEAL~\cite{arabi2025seal} & 0.999 & 0.036 & 0.068 & 0.090 & 0.060 & 0.050 & $7.1{\times}$/$5.8{\times}$ \\
    \bottomrule
  \end{tabular}}
\end{table}

\subsection{Image Quality Evaluation}
\label{sec:quality}


Table~\ref{tab:quality_main} evaluates perceptual impact. The codec embedding yields PSNR $26.81{\pm}2.69$ dB, SSIM $0.811{\pm}0.069$, and LPIPS $0.221{\pm}0.032$, while the paired GS-vs-Dual comparison remains close to plain-vs-codec (LPIPS $0.220{\pm}0.031$), suggesting that the GS branch contributes only a small incremental distortion relative to the full Dual-Guard output. CLIP-T declines by only 1.1\% from plain (0.327) to Dual-Guard (0.323).

\begin{table}[t]
  \caption{Image quality evaluation on 1,000 paired images. Upper rows report paired distortion metrics including LPIPS. Lower rows report CLIP-T semantic alignment per variant.}
  \label{tab:quality_main}
  \centering
  \resizebox{\columnwidth}{!}{
  \begin{tabular}{lccccccc}
    \toprule
    Comparison & PSNR$\uparrow$ & SSIM$\uparrow$ & LPIPS$\downarrow$ & L1$\downarrow$ & CLIP-I$\uparrow$ & CLIP-T$\uparrow$ & $\Delta$CLIP-T \\
    \midrule
    plain-vs-codec & $26.81{\pm}2.69$ & $0.811{\pm}0.069$ & $0.221{\pm}0.032$ & $0.031{\pm}0.010$ & $0.966{\pm}0.017$ & -- & -- \\
    GS-vs-Dual     & $26.78{\pm}2.70$ & $0.813{\pm}0.068$ & $0.220{\pm}0.031$ & $0.031{\pm}0.010$ & $0.966{\pm}0.019$ & -- & -- \\
    \midrule
    plain       & -- & -- & -- & -- & -- & $0.327{\pm}0.035$ & -- \\
    GS-only     & -- & -- & -- & -- & -- & $0.326{\pm}0.033$ & $-$0.001 \\
    codec-only  & -- & -- & -- & -- & -- & $0.324{\pm}0.036$ & $-$0.003 \\
    Dual-Guard  & -- & -- & -- & -- & -- & $0.323{\pm}0.035$ & $-$0.004 \\
    \bottomrule
  \end{tabular}}
\end{table}

\subsection{External Baseline Comparison}
\label{sec:baseline_reruns}


Table~\ref{tab:baseline_reruns} reruns three diffusion-watermark baselines under the same prompt snapshot. SEAL~\cite{arabi2025seal} is reported with its released evaluation path and native patch-L2 / spatial-test protocol, without retraining on our 1,000-prompt benchmark. Blank cells denote metrics not directly exposed by the corresponding stable upstream evaluation path under our matched setup. This table is therefore a task-aligned comparison against diffusion-watermark baselines rather than an exhaustive benchmark against classical forensic localizers. Under this protocol, GS and Tree-Ring retain provenance but no localization, SEAL localizes edits at only 0.036/0.068 IoU/F1, and Dual-Guard is the only method here that both localizes and detects $\geq$99.9\% of reprompt attacks.

\begin{table}[t]
  \caption{External baseline comparison under the shared 1,000-prompt snapshot. SEAL uses its native patch-L2 / spatial-test protocol.}
  \label{tab:baseline_reruns}
  \centering
  \resizebox{\columnwidth}{!}{
  \begin{tabular}{lcccccc}
    \toprule
    Method & \shortstack{Reprompt\\Defense} & \shortstack{Clean\\AUC} & \shortstack{Attack\\AUC} & \shortstack{Loc.\\IoU} & \shortstack{Loc.\\F1} & \shortstack{Loc.\\Support} \\
    \midrule
    Gaussian Shading~\cite{yang2024gaussianshading} & \texttimes & 1.000 & 0.998 & --    & --    & \texttimes \\
    Tree-Ring~\cite{wen2024treering}                & \texttimes & 1.000 & 0.998 & --    & --    & \texttimes \\
    SEAL~\cite{arabi2025seal}                       & \texttimes & 1.000 & --    & 0.036 & 0.068 & \checkmark \\
    \midrule
    Dual-Guard       & \checkmark & 1.000 & 1.000 & \textbf{0.255} & \textbf{0.392} & \checkmark \\
    \bottomrule
\end{tabular}}
\end{table}

This comparison clarifies the division of labor across methods. GS- and Tree-Ring-style schemes remain strong model-of-origin markers, but by themselves they cannot tell whether the current content still matches the claimed issuance record after regeneration. SEAL contributes spatial evidence, yet without a claim-binding provenance anchor it does not address reprompted framing. Dual-Guard closes both gaps by pairing a robust provenance key with a record-specific latent reference.

\subsection{Ablation Study}
\label{sec:ablation}

\noindent\textbf{Reference Latent and Calibration Consistency.}
Table~\ref{tab:ablation_ref} shows that round-trip reference construction and matched calibration are essential: the generated-reference stress test drives clean pass to 0.000, highlighting sensitivity to calibration mismatch rather than invariance across arbitrary reference-construction pipelines.

In deployment terms, $z_0^{\text{ref}}$ should therefore be treated as an owner-side record artifact created at issuance time, not as a latent that can be regenerated interchangeably from the prompt or from a different reconstruction path. The verifier is calibrated against VAE round-trip references, so changing that construction shifts the score distribution seen by the block comparison stage.

\begin{table}[t]
  \caption{Stress test on reference-latent construction and calibration consistency. The generated-reference variant reuses round-trip-calibrated thresholds and is included specifically to expose calibration mismatch, not as a deployment recommendation.}
  \label{tab:ablation_ref}
  \centering
  \resizebox{\columnwidth}{!}{
  \begin{tabular}{lccccc}
    \toprule
    Variant & \shortstack{Clean\\Auth Pass $\uparrow$} & \shortstack{False Tamper\\Alarm $\downarrow$} & \shortstack{Reprompt\\Detect $\uparrow$} & \shortstack{Local\\Detect $\uparrow$} & \shortstack{DiffEdit\\Detect $\uparrow$} \\
    \midrule
    Generated-ref        & 0.000 & 1.000 & 1.000 & 1.000 & 1.000 \\
    Round-trip (default) & 0.997 & 0.001 & 1.000 & 0.999 & 1.000 \\
    \bottomrule
  \end{tabular}}
\end{table}

\noindent\textbf{Dual-Branch Complementarity.}
Table~\ref{tab:ablation_branch} shows that GS-only misses 97.5\% of local attacks and 60.1\% of DiffEdit cases, whereas full Dual-Guard reaches $\geq$99.9\% across all scenarios.

This split also explains the decision order in Full mode. GS first answers whether the claim points to a protected issuance path at all, while the codec branch then asks whether the current image still matches that specific record. Localization is meaningful only after those two checks agree; otherwise the correct outcome is claim rejection or content mismatch rather than region marking.

\begin{center}
\begin{minipage}{\columnwidth}
  \captionsetup{type=table}
  \captionof{table}{Ablation on channel contributions (2,400-sample benchmark). Localization gain measures the fraction of attacks caught exclusively by the localization branch.}
  \label{tab:ablation_branch}
  \centering
  \resizebox{\columnwidth}{!}{
  \begin{tabular}{lcccccc}
    \toprule
    Configuration & \shortstack{Reprompt\\Detect $\uparrow$} & \shortstack{Local\\Detect $\uparrow$} & \shortstack{DiffEdit\\Detect $\uparrow$} & \shortstack{Local\\Missed $\downarrow$} & \shortstack{DiffEdit\\Missed $\downarrow$} & \shortstack{Loc.\\Gain} \\
    \midrule
    GS + global gate only & 1.000 & 0.025 & 0.399 & 97.5\% & 60.1\% & -- \\
    Full Dual-Guard       & 1.000 & 0.999 & 1.000 & 0.1\% & 0.0\% & 0.601--0.975 \\
    \bottomrule
  \end{tabular}}
\end{minipage}
\end{center}

\section{Conclusion}
\label{sec:conclusion}

\begin{sloppypar}
We presented Dual-Guard, a dual-channel latent watermarking framework for provenance verification, reprompting defense, and region-level tamper localization in diffusion-generated images. In the platform-side closed-set setting, Full mode combines GS provenance checks with reference-latent block verification, achieving a 99.7\% clean authentication pass rate, a 0.1\% clean tamper false-alarm rate, and ${\geq}$99.9\% detection for reprompting, local tampering, and DiffEdit, with block-level IoU~0.255 and F1~0.392. The results suggest that provenance verification and tamper localization are most reliable when a robust model-of-origin signal is explicitly tied to a record-specific content anchor instead of being asked to do both jobs alone. This deployment point is especially natural for creator platforms or asset registries that can retain owner-side records, while lighter fingerprint-only operation remains available when the round-trip reference cannot be retrieved. Future work includes broader forensic comparisons, open-set fingerprint search, stronger white-box adaptive attacks, and lower image-quality cost.
\end{sloppypar}

\bibliographystyle{plain}
\bibliography{references}

\end{document}